\documentclass[conference]{IEEEtran}
\usepackage{graphicx} 
\usepackage{amsmath}
\usepackage{amsfonts}
\usepackage{hyperref}
\usepackage{xcolor}

\usepackage{amsmath,pdftexcmds}

\DeclareFontFamily{U}{cbgreek}{}
\DeclareFontShape{U}{cbgreek}{m}{n}{
        <-6>    grmn0500
        <6-7>   grmn0600
        <7-8>   grmn0700
        <8-9>   grmn0800
        <9-10>  grmn0900
        <10-12> grmn1000
        <12-17> grmn1200
        <17->   grmn1728
      }{}
\DeclareFontShape{U}{cbgreek}{bx}{n}{
        <-6>    grxn0500
        <6-7>   grxn0600
        <7-8>   grxn0700
        <8-9>   grxn0800
        <9-10>  grxn0900
        <10-12> grxn1000
        <12-17> grxn1200
        <17->   grxn1728
      }{}

\DeclareRobustCommand{\Qoppa}{%
  \text{\usefont{U}{cbgreek}{\normalorbold}{n}\symbol{21}}%
}
\makeatletter
\newcommand{\normalorbold}{%
  \ifnum\pdf@strcmp{\math@version}{bold}=\z@ bx\else m\fi
}
\makeatother

\title{On Using The Path Integral Formalism to Interpret Synchronization in Quantum Graph Networks}
\author{\IEEEauthorblockN{JTM Campbell}
\IEEEauthorblockA{Department of Electronic Engineering\\
Maynooth University\\
Ireland\\
Email: ektopyrotic@gmail.com}}
\date{August 2024}

\begin{document}

\maketitle

\begin{abstract}
This article explores the application of the path integral formalism in describing synchronization phenomena in entangled networks, cavities, and reservoirs. We discuss the concept of using Lagrangian mechanics for systems undergoing synchronization and its connection to least-action principles. By replacing the concept of least action with a least signaling term, we investigate how the path integral representation can be applied to study synchronization dynamics in entangled networks, drawing parallels with coupled oscillators in phase space models such as the Kuramoto model, as well as its relation to algorithms, such as the firefly algorithm for potential use in optimization in networks. This article also illustrates how entanglement signals themselves can interact strongly with ordered systems of harmonic oscillators that reach thresholds of classical synchronization with potential therefore for using entangled signals as weak measurement probes where phase dynamics is of interest. 
\end{abstract}

\section{Introduction}
The concept of synchronization has a rich history, with Christian Huygens' discovery of synchronized oscillations in pendulum clocks serving as a seminal example [1]. Work done by Rayleigh on phasor sums of independent signals has since literally shaped the statistical analysis of synchronized signaling systems.[2] The circular statistics of Rayleigh was extended by work pioneered by Kuramoto and others [3], relating synchronization with phase-space descriptions of systems of oscillators that produce models that can be extended to describe various emergent phenomena in natural and engineered systems [4][5]. Meanwhile, the description of quantum mechanics in phase space by Wigner and Husimi allowed work to extend the Feynman Path Integral formalism in Quantum Mechanics to phase space [6], forming a critical bridge between the description of many-body quantum systems and their evolving trajectories in complex entangled systems. In this article, we explore the use of the path integral formalism to describe synchronization in signaling entangled networks. By connecting the Lagrangian formalism to network synchronization and extending it to phase space models like the Kuramoto model, we aim to shed light on the intricate dynamics of synchronized systems, with an emphasis on how entanglement can be related to percolation thresholds where synchronization/entanglement can occur, and how at an "edge of percolation," a quantum channel may enter a "chimera state" of entanglement, i.e., a so-called discordant state.

\section{Lagrangian and Least Signaling Terms}
In the context of network synchronization, the Lagrangian formalism can be extended to describe the dynamics of a network of coupled oscillators. The Lagrangian for a network of \( N \) coupled oscillators can be expressed as:
\begin{equation}
\mathcal{L} = \sum_{i=1}^{N} \frac{1}{2} m_i \dot{q}_i^2 - V(q_1, q_2, \ldots, q_N),
\end{equation}
where \( m_i \) is the mass of oscillator \( i \), \( \dot{q}_i \) is the velocity of oscillator \( i \), and \( V \) represents the potential energy of the coupled oscillators.

The action \( S \) for the network of coupled oscillators is given by:
\begin{equation}
S = \int_{t_1}^{t_2} \mathcal{L} dt,
\end{equation}
where \( t_1 \) and \( t_2 \) denote the initial and final times, respectively.

By replacing the action \( S \) with a least signaling term \( P_{\text{send}} \), we capture the minimal signaling requirements necessary for synchronization to occur in entangled networks. This least signaling term is defined mathematically as:
\begin{equation}
P_{\text{send}} = \int_{t_1}^{t_2} dt \ \mathcal{L}_{\text{send}}(q_i, \dot{q}_i),
\end{equation}
where \( \mathcal{L}_{\text{send}} \) represents the signaling dynamics with the signaling connections among the oscillators and may depend on the state variables and their derivatives over time. This term encapsulates the essence of information exchange necessary for achieving synchronization.

\subsection{Path Integral Representation}
In the path integral formulation of quantum mechanics, a system can take any trajectory between two points of observation \( x_a \) and \( x_b \), at times \( t_a \) and \( t_b \), respectively. Denoting a path as \( x(t) \) such that \( x(t_a) = x_a \) and \( x(t_b) = x_b \), the probability amplitude of the system taking this path is given by:
\begin{equation}
A[x(t)] \propto e^{-\frac{i}{\hbar} S[x(t)]},
\end{equation}
where the action \( S[x(t)] \) is defined as:
\begin{equation}
S[x(t)] = \int_{t_a}^{t_b} dt \ L(\dot{x}(t), x(t)).
\end{equation}
Therefore, the overall probability amplitude of the system starting at \( x_a \) at time \( t_a \) and ending up at \( x_b \) at time \( t_b \) is given by the sum over all possible paths:
\begin{equation}
\psi(x_b, t_b; x_a, t_a) \ \propto \sum_{x(t)} e^{-\frac{i}{\hbar} S[x(t)]}.
\end{equation}
While it is challenging to rigorously define a sum over all paths, we can approximate paths \( x(t) \) as continuous variables, treating the action \( S[x(t)] \) as a function of this variable.

Although the propagator, \( \psi(x_b, t_b; x_a, t_a) \) does not directly encapsulate effective coupling in oscillator networks, there exists an analogous model we can consider that can elucidate some of the emergent dynamics of synchronized states and collective oscillations represented in graph theory that can be bridged to represent dynamics in highly entangled networks [4], [7].

\section{Kuramoto Model in Phase Space}
The Kuramoto model provides a powerful framework for studying the synchronization of coupled oscillators in phase space. The dynamics of the Kuramoto model can be described by the following equation:
\begin{equation}
\dot{\theta}_i = \omega_i + \sum_{j=1}^{N} X_{ij} \sin(\theta_j - \theta_i),
\end{equation}
where \( \theta_i \) represents the phase of oscillator \( i \), \( \omega_i \) is the natural frequency of oscillator \( i \), \( X_{ij} \) denotes the coupling strength between oscillators \( i \) and \( j \), and \( N \) is the total number of oscillators in the network.

The Kuramoto model captures the collective behavior of coupled oscillators and how synchronization emerges through the interaction of phase differences between oscillators. By analyzing the phase space trajectories of the oscillators, one can observe the evolution of synchronization patterns and the emergence of coherent states in the network.

\subsection{Effective Behaviors in a Network of Oscillators}
The effective behaviors of oscillators within a network can be comprehensively analyzed through the lens of graph theory, with mathematical insights that draw from quantum mechanics, particularly regarding coherent states and entangled systems. The effective coupling strength between two oscillators in such a network can be articulated through their coordinated oscillation patterns represented on the graph.

\subsubsection{Graph Representation}
Consider an oscillator network represented as a graph \( G = (V, E) \) where \( V \) denotes the set of vertices corresponding to oscillators (or modes), and \( E \) encapsulates the edges that define the coupling strengths \( X_{ij} \) connecting oscillators \( i \) and \( j \). Each edge weight is indicative of the interaction strength between adjacent oscillators, thus affecting their collective dynamics [5].

\subsubsection{Effective Coupling}
The effective coupling strength \( X_{\text{eff}}(i, j) \) between oscillators \( i \) and \( j \) can be determined using various methods, such as examining mode shapes or invoking Hamiltonian dynamics. A robust analytical method is to analyze the synchronization properties of the network. For instance, in a coupled oscillator model, one may express the effective coupling as:
\begin{equation}
X_{\text{eff}}(i, j) = \sum_{k \in N(i)} X_{ik} C_k,
\end{equation}
where \( N(i) \) is the set of neighbors of oscillator \( i \), and \( C_k \) represents the contributions of these neighboring states to the effective coupling.

\subsubsection{Connection to Synchronized States}
In a synchronized oscillator system, the dynamics can be characterized by mutual synchronization, wherein the collective oscillations arise from the intricate coupling in the network. The propagator \( \psi(x_b, t_b; x_a, t_a) \) encapsulates the amplitudes for transitions between coherent states of oscillators \( i \) and \( j \). Drawing similarities with the Feynman picture of quantum sum-over-histories [6] The synchronization propagator is also an effective manifestation of collective oscillations, which in a network of oscillators will stem from the underlying graph structure.

\subsubsection{Path Integral for Synchronization}
Interpreting the effective coupling through the framework of path integrals lends further insight: we can view the effective coupling as arising from the cumulative effects of all possible paths linking oscillatory states. Each path’s contribution is weighted by the respective coupling strengths along the paths:
\begin{equation}
X_{\text{eff}}(x_a, x_b) = \sum_{\text{paths}} W(\text{path}) \cdot \text{Amplitude},
\end{equation}
where \( W(\text{path}) \) represents the weight associated with the coupling strength along a given oscillatory path.

\subsubsection{Mathematical Formalism}
By employing graph theory in concert with the principles of oscillator dynamics, the effective coupling can be expressed in terms of the adjacency matrix \( A \) of the graph or through the eigenfunctions of the Hamiltonian associated with the system. Specifically, it can be represented as:
\begin{equation}
X_{\text{eff}}(x_a, x_b) = \frac{A_a}{B},
\end{equation}
where \( A_a \) represents the amplitude of the oscillators at positions \( a \) and \( b \), and \( B \) is a measure of collective coupling strength across the involved oscillators.

The Kuramoto model provides a phase space description of coupled oscillators, where the dynamics of synchronization are governed by the phase differences between oscillators. In this model, the coupling strength between oscillators influences the emergence of synchronized states in the network.

\section{Discontinuous Coupling in Phase Space}
Let us take our oscillators running on a unit circle in phase-space. The oscillators represented on a circle in phase space. To achieve synchronization, the oscillators runs clockwise on the circle until it passes the threshold. Whenever it passes the threshold, it will emit a signal pulse, with some probability, \( P_{\text{send}} \). When the oscillators are coupled, they will adjust their individual rates, or periods, of signaling to match with one another (encroaching closer together on the unit circle in phase-space) to achieve synchronization under a time evolution.

\begin{figure}[ht]
    \centering
    \includegraphics[width=0.45\textwidth]{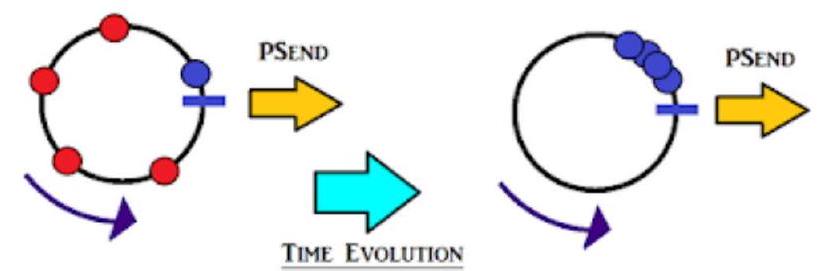}
    \caption{Oscillators running on a unit circle in phase space, emitting signal pulses when passing threshold.}
    \label{fig:unit_circle_pulses}
\end{figure}

Each oscillator will adjust its period of signaling according to our signaling activation function, which is, under the signaling field representation, a phase function:
$$
\Phi\left(X_{i, j}\right)=A\left(X_{i, j}\right) \cdot \alpha
$$
The function evolves linearly over time until it reaches a threshold value, defined by the vector weights, \( \mathbf{X} \mathbf{i}, \mathbf{j} \), which are translated into coupling strengths under a power law, i.e. the extinction ratio for light signals.

When the threshold is reached, a single oscillator fires the signal pulse and then resets its phase. If no coupling occurs, the oscillator will pulse with a period, \( T(A) \). \( T \) is effectively the encoded time remaining until the next signal pulse.

\begin{figure}[ht]
    \centering
    \includegraphics[width=0.50\textwidth]{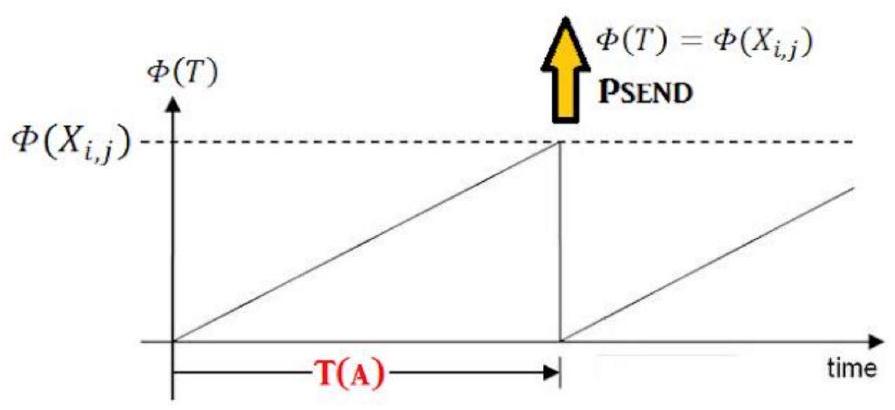}
    \caption{Signaling activation function at threshold.}
    \label{fig:activation_function}
\end{figure}

When a phase-coupling occurs between oscillators, when an oscillator receives a pulse it will increment its phase function by an amount that itself depends on its current value and the change induced by the weights of the received signal. The oscillator will then pulse with a new period, \( T(B) \):

\begin{figure}[ht]
    \centering
    \includegraphics[width=0.50\textwidth]{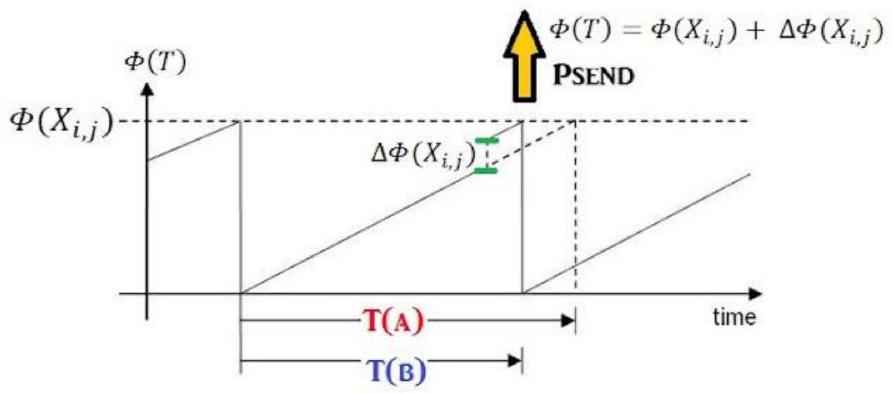}
    \caption{Phase updates in response to incoming signals.}
    \label{fig:phase_updates}
\end{figure}

This means that an oscillator that is in the lower half of the phase-space circle will jump forward, and an oscillator that is in the upper half will jump backwards in the circle in phase space. This is the main concept of a self-organizing update function, which brings the 2 oscillators in phase with one another and thus coupled.

The resemblance of the unitary time-evolution operator in quantum mechanics to the monotonically decreasing signaling function is obvious in their respective mathematical forms:
$$
\beta=\beta_{0} e^{-\gamma r^{m}} \text{ and } U=U_{0} e^{\frac{-i H t}{\hbar}}
$$

Moreover, like the unitary time-evolution operator, the signaling algorithm is readily compatible with Lagrangian mechanics in the same way as the unitary time-evolution operator which are inherently used for discrete particles each with a finite number of degrees of freedom.

The classic Lagrangian is written as the difference of the kinetic and potential energy density of a particle.
$$
L=K \cdot E - P \cdot E = \frac{m v^{2}}{2}-V(q)
$$

The potential energy density of the particle is a field density, such as an electromagnetic field density in generalised coordinates \( V(q) \).

Since \( v=p / m \), the kinetic energy can, of course, be written in terms of momentum, \( p \), which is more relevant for what we are describing.
$$
L=\frac{p^{2}}{2 m}-V(q), \quad p=\frac{d L}{d q}
$$

The abbreviated action \( \mathbf{S_{0}} \) is defined as the integral of the generalized momenta along a path in the generalized coordinates \( q \):
$$
S_{0}=\int p \cdot d q=\int p_{i} d q_{i}
$$

In our ad-hoc system of synchronization coupling, the momenta will be replaced by \( P_{\text{send}} \) and integrated over the time domain (dt) in the path it takes to achieve synchronization. This leads to the total Signaling Action, \( \mathbf{S_{f}} \), being represented as the phase of each path being determined by \( \int P_{\text{send}} \mathbf{d t} \), for that trajectory.
$$
S_{f}=\int P_{\text{send}} d t
$$

The frequency of signaling is then, as an oscillation in phase-space:
$$
e^{\boldsymbol{S}_{\boldsymbol{f}}}
$$

The frequency of signaling is then an oscillation in the complex plane. In the nature of the path integral formalism, we consider the synchronization as:
\[
\left\langle \textcolor{blue}{\boldsymbol{\bigcirc}} \left| \beta_0 e^{-\gamma t} \right| \textcolor{red}{\boldsymbol{\bigcirc}} \right\rangle
\]

which can be visualized as a sum of signaling histories as:

\begin{figure}[h]
    \centering
    \includegraphics[width=0.50\textwidth]{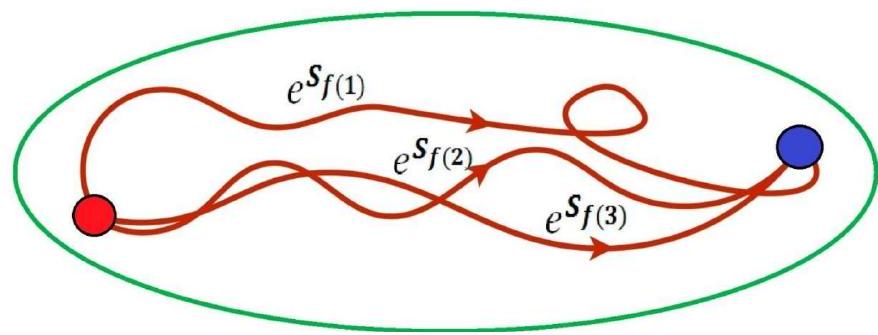}
    \caption{\textcolor{teal}{\textbf{Configuration space}} containing the total amplitude for synchronization history; it encapsulates the cumulative effects of all allowed signals, each influenced by their respective signaling actions.}
    \label{fig:sync_history}
\end{figure}

In this convention, the signaling action, \( \mathbf{S_{f}} \), which is a real operator is defined as being essentially characteristic of the physical and environmental parameters of the system - namely the light absorption coefficient and the power law over the distance by which it is subject to.

\section{Quantum Network Simulations using Kuramoto Oscillators}

The path integral formalism hence offers a comprehensive framework for studying the dynamics of entangled networks undergoing synchronization. By representing the least signaling action in configuration space as a sum over functions, \( \sum e^{P_{\text{send}}} \), we can extend this representation to phase space models akin to the Kuramoto model, from which we can develop several pseudo-entanglement networks to run simulations. The phase space description enables us to analyze the collective behavior of coupled oscillators in our models and understand the mechanisms underlying synchronization and discordance in entangled networks.

To do this with a comprehensive scope a series of simulation codes are used in python building on the complex systems work of Hiroki Sayama et al  [10].

The results of using Kuramoto oscillators in phase space is showcased here: https://github.com/MuonRay/QuantumNetworkSimulations

these series of codes use networkx in python to create network graphs (Fig 5) with nodes and links subject to the description in the section III-A on Graph Representation.

\begin{figure}[ht]
    \centering
    \includegraphics[width=0.5\textwidth]{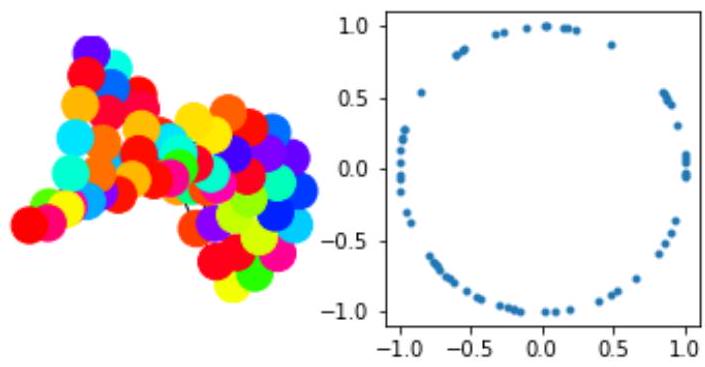}
    \caption{Left: graph G = (V, E) where V denotes the
set of vertices corresponding to oscillators (or modes), and E
encapsulates the edges that define the coupling strengths Xij
connecting oscillators i and j. Right: Phase space distribution of oscillators
}
    \label{fig:unit_circle_pulses}
\end{figure}

From this a series of experiments were undertaken to drive a system of synchronized oscillators and from using the network path-integral formalism discussed a process was developed to get to an action-based descriptor of the networks examined as they are driven towards synchronization from a initial condition of chaos towards an underdamped state, a critically damped threshold and ultimately a steady-state. The description of this network of oscillators as an underdamped system can be seen when the derivative of the action is plotted (Fig 6)

\begin{figure}[ht]
    \centering
    \includegraphics[width=0.5\textwidth]{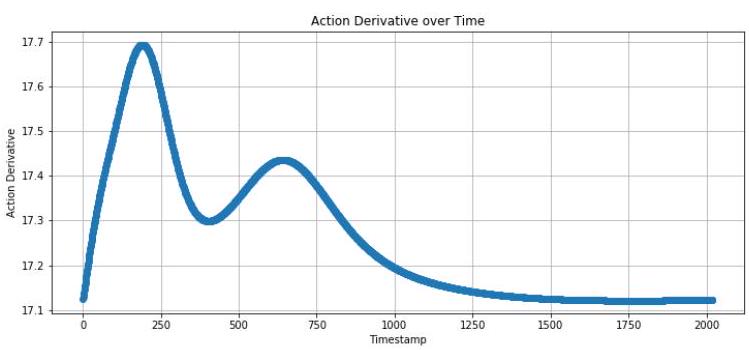}
    \caption{derivative of the action, S, plotted over the time of the evolution of synchronization, showing the underdamped nature of the system as well as the critical points of change
}
    \label{fig:unit_circle_pulses}
\end{figure}

The evolution of this system of oscillators can be compared in the degree of chaotic phases of the oscillators, which relates to the frequency shifts per timestep (Fig 7). 

\begin{figure}[ht]
    \centering
    \includegraphics[width=0.5\textwidth]{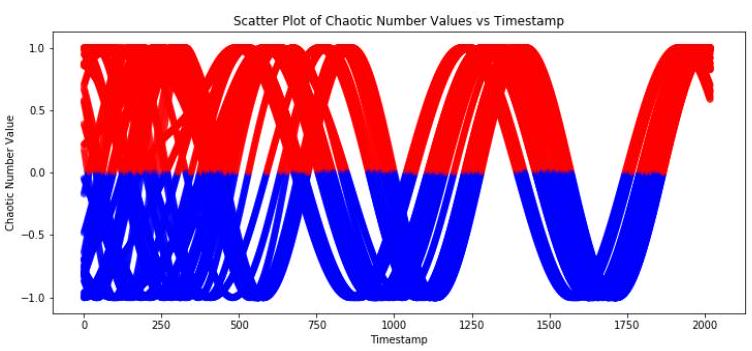}
    \caption{chaotic phases of the oscillators in an underdamped network approaching synchronization in a time evolution
}
    \label{fig:unit_circle_pulses}
\end{figure}

It is possible combine these factors, which show relatable common trends, to give an overall configuration space trajectory of the system, where a colorbar can show exactly the critical point at which the maximum and minima actions drive the networked system of oscillators from underdamped to a steady state as they synchronize (Fig 8).

\begin{figure}[ht]
    \centering
    \includegraphics[width=0.50\textwidth]{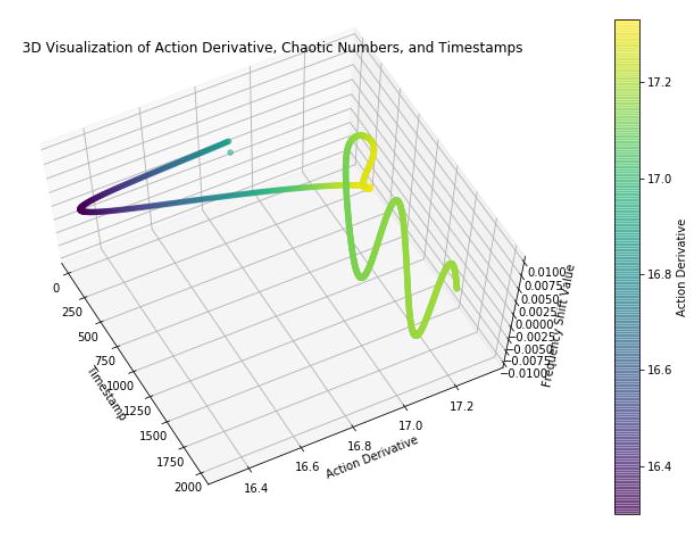}
    \caption{Trajectory of the Network of Oscillators in a high dimensional configuration space as synchronization emerges. It is possible for the trajectory to intersect itself, because the phase configuration of the system can go back to the same position after a given interval of time.}
    \label{fig:unit_circle_pulses}
\end{figure}

The motivation for these simulations is that a coupled system of harmonic oscillators, such as an entanglement network, will need to be examined under the lens of control theory used in classical complex systems. For a salient description of how these entangled networks interact with the environment models must be considered that view them in terms familiar to analog systems of coupled oscillators that behave subject to perturbations, i.e. sudden changes or disturbances in the system, whether in the network parameters, external forces, or the configuration of the nodes/links in the network. 

In networked structures, the coupling between different nodes means that a perturbation in one part of the system can influence others. Therefore, the response of the system (in terms of the action's derivative) can exhibit complex dynamics depending on the network topology and interactions.

The interconnectivity allows perturbations that do not just affect one node to propagate through the network, manifesting in collective phenomena such as synchronization, cascading failures, or robustness to disturbance.

\section{Entangled Signal Interaction with Classical Synchronization Networks}
One intriguing aspect of this interaction between entanglement and synchronization phenomena in networks of coupled oscillators is how entangled signals can serve as probes in classical synchronization scenarios. Entangled states, characterized by correlations that cannot be explained classically, carry unique phase properties that can influence the behavior of systems described as networks of harmonic oscillators.

As harmonic oscillators within a network approach thresholds of synchronization, the stability and coherence of their collective dynamics can be significantly affected by the presence of entangled signals. When entangled signals interact with these oscillators, they can facilitate synchronization under certain conditions, effectively acting as a catalyst for the emergence of synchronous behavior. This interaction is particularly relevant at what is termed the "edge of synchronization," where small perturbations—such as those introduced by quantum entanglement—may shift the system into a regime of enhanced synchrony.

Moreover, the interplay between entangled signals and harmonic oscillators presents exciting possibilities for weak measurement probes, where the measurement process minimally disturbs the system being observed. [8] In scenarios where phase dynamics are of primary interest, entangled signals can be deployed to extract phase information without fully collapsing the wave function of the oscillator network. This capability opens up avenues for leveraging entangled states to enhance our understanding of phase relationships and synchronization dynamics within quantum systems.

In the context of a network of oscillators, a weak measurement operator, denoted as \( A_\omega \), acts on the eigenmodes of the system, offering insights into its phase dynamics. By utilizing weak values \( A_w \), one can investigate how oscillators interact, particularly in entangled states, without collapsing the wavefunction to a definitive outcome. This capacity allows the exploration of the intricate nature of their synchronizing relationships, reminiscent of the synchrony observed in classical oscillators, albeit with the added complexity of non-local entanglement behavior.

When weak measurement operators are reformulated within the framework of the path integral approach, they can quantify the probable trajectories that oscillators might take during synchronization. Each trajectory can be interpreted as a quantum path, weighted by its action \( S \):
\[
\mathcal{Z} = \int \mathcal{D}[r(t)] e^{\frac{i}{\hbar} S[r(t)]}
\]

By measuring the dynamics of each state using weak measurement operators, one can establish a connection between quantum measurements and classical paths in configuration space.

The resemblance of the time-dependent weak measurement operator to the monotonically decreasing signaling function is obvious in their respective mathematical forms:
\begin{equation}
    \hat{M}(t) = \int_{-\infty}^{t} K(t, s) \hat{O}(s) \, ds,
\end{equation}
where \( \hat{M}(t) \) denotes the weak measurement operator at time \( t \), \( \hat{O}(s) \) represents the observable being measured at time \( s \), and \( K(t, s) \) is the kernel function that defines the influence of the past measurements on the present measurement outcome.

The eigenmodes can be quantified as variables measured in a weak measurement operator, \( \hat{M}(t) \). 

The operator \( \hat{M}(t) \) then acts on the eigenmodes, with the initial and final states of the propagator \( \psi(x_f, t_f; x_i, t_i) \):  

\begin{equation}
\hat{M}(t) = \frac{\langle \psi_{f} \rvert \hat{O}(s) \lvert \psi_{i} \rangle}{\langle \psi_{f} \rvert \psi_{i} \rangle}
\end{equation}
with the observable operation \( \hat{O}(s) \).

In comparison, the monotonically decreasing signaling function, \( f(t) \), can be expressed as:
\begin{equation}
    f(t) = \beta_0 e^{-\gamma t},
\end{equation}
where \( \beta_0 \) is the initial signal strength and \( \gamma \) is a positive constant that dictates the rate of decay.

Both expressions illustrate a dependency on past values and exhibit a form of time evolution characterized by an integral or exponential decay, highlighting their analogous behavior in temporal dynamics.

This is readily compatible with Lagrangian mechanics which are inherently used for discrete particles each with a finite number of degrees of freedom. Our Signalling Lagrangian, \( L_f \), for each signaling node is in effect integrated across the time-domain in the path it takes to achieve synchronization, hence leading to the total Signalling Action \( S_f \), represented here as being the phase of each path being determined by \( \int L_f \mathbf{d r} \), for that trajectory, where \( \mathrm{L_f} \) is the Lagrangian.
$$
S_{f}=\int L_{f} d t=\ln \left(\frac{I(t)}{I_{0}}\right)=-\gamma t
$$

In this convention, the signaling action, \( S_f \), which is a real operator is defined as being essentially characteristic of the physical and environmental parameters of the system - namely the light absorption coefficient and the power law over the distance by which it is subject to.

Since each deviation from the path of least action is, just like in the case with a quantum particle, proportional to the action imparted on the system,

the signal activation function will activate with a frequency proportional to the action.
Therefore, the path of least possible action, which occurs during complete internal synchronization, should have the lowest frequency of the signaling function.

\vspace{10mm}

\subsection{Linking Deviation in Frequency to Action}

In weak measurement theory, we consider a system (such as a photon with spin states) that is slightly perturbed by an interaction with a measurement device. The initial state of the system can be written as:
\[
|\psi\rangle = \alpha |s_1\rangle + \beta |s_2\rangle
\]
where \( |s_1\rangle \) and \( |s_2\rangle \) are the eigenstates corresponding to spin up and spin down, respectively, and \( \alpha \) and \( \beta \) are complex amplitudes with the normalization condition \( |\alpha|^2 + |\beta|^2 = 1 \).

The interaction Hamiltonian \( H' \) can be represented as:
\[
H' = -\hbar \Delta \omega \hat{P}
\]
where \( \hat{P} \) is the operator corresponding to the observable being measured (here, it may relate to the spin projection), and \( \Delta \omega \) represents the perturbation in frequency due to the measurement process.

The energy associated with this perturbation is:
\[
\Delta E \approx \hbar \Delta \omega
\]

The integration of weak measurement techniques within this framework allows for the observation of these subtle variations in frequency without significantly disturbing the system. [9] Under this paradigm, we can effectively monitor synchronization integrity through measurable frequency shifts related to the action’s time derivative .

When a synchronized system experiences a fault, it deviates from the path of least action. This deviation can be quantified in terms of a change in frequency \( \Delta \omega \):
\[
\Delta \omega \propto \frac{\partial S}{\partial t} \bigg|_{\text{faults}}.
\]
This implies that as the action changes due to faults in the synchronization, the frequency of the system’s oscillations will also change. 

The proportionality indicates that greater deviations from the optimal path correspond to more significant changes in frequency [6].

\[
\Delta \omega \propto \Qoppa\frac{\partial S}{\partial t}
\]

This notion allows for the monitoring of synchronization integrity through observable frequency shifts with respect to a drive factor \( \Qoppa \).

Thus, the weak measurement framework, when aligned with the path integral formalism, affords novel insights into the delicate synchronization dynamics of both classical and quantum systems. This approach serves as a mechanism for testing theories of synchronization across both domains, leveraging weak values to elucidate the underlying structures and dynamics governing synchronized states.

The potential applications of this work are broad, ranging from weak measurement approaches to quantum information to studies of biological systems exhibiting synchronized behavior. As such, understanding the dynamics of entangled signals within ordered systems of harmonic oscillators paves the way for innovative approaches to control and measure both classical and quantum complex systems.

\section{Applications and Future Directions}
The application of the path integral formalism to investigate synchronization in entangled networks presents exciting opportunities for advancing our understanding of complex synchronization phenomena. Future research directions could explore the quantum aspects of network synchronization, develop computational tools based on path integral methods for network analysis, and apply these concepts to real-world networked systems for enhanced synchronization control and measurement.

\section{Conclusions}
In this article, we have explored the utilization of the path integral formalism to describe synchronization and percolation phenomena in entangled networks. By linking the Lagrangian formalism to network synchronization dynamics and extending it to phase space models like the Kuramoto model, we have uncovered new insights into the synchronization behavior of complex systems in a quantum context. This can lead to the development of new optimization techniques to purify quantum states for distribution and measurement in scalable networks and for leveraging synchronization principles with control theory.

\subsection*{Appendix: Frequency Shift Relation to Action deviations}

\subsection*{Express Changes in Energy:}
Combining the previous steps, we can express changes in energy in terms of the action:
\[
\Delta E \approx \hbar \left( \Qoppa \frac{\partial S}{\partial t} \right)
\]
where \(\Qoppa\) is a proportionality constant that depend on the oscillator network system under consideration, i.e. our network of synchronized oscillators.

\subsection*{Energy and Wavelength:}
The relationship between energy and wavelength can be expressed in two equivalent forms:
\[
E = \frac{h}{\lambda} \quad \text{or} \quad E = \hbar k, \quad k = \frac{2\pi}{\lambda}
\]
From these relationships, we can deduce:
\[
\Delta E \approx h \Delta \left( \frac{1}{\lambda} \right)
\]

\subsection*{Equating Energy Expressions:}
By equating the different expressions for energy, we have:
\[
\hbar \left( \Qoppa \frac{\partial S}{\partial t} \right) = h \Delta \left( \frac{1}{\lambda} \right)
\]

\subsection*{Wavelength Shift Formula:}
From the previous equation, we can derive the wavelength shift formula:
\[
\Delta \lambda \propto -\frac{\hbar}{h} \left( \Qoppa \frac{\partial S}{\partial t} \right)^{-1}
\]
This indicates that the change in wavelength is inversely related to the changes in action. it also allows us to work with a network fault constant $\Qoppa $ that works on the rate of change to the action on our oscillator network.

If we wish To remove the inverse we then establish a relation between the frequency shift and the partial derivative of the action with respect to time:
\[
\Delta \omega \propto \Qoppa\frac{\partial S}{\partial t}
\]
This implies that the change in frequency is associated with the rate of change of the action.

This relationship suggests that changes in wavelength can originate from alterations in the underlying physical processes encapsulated by the action.

\vspace{135mm}
\thanks{[This material is based upon work supported by Science Foundation Ireland (SFI) and is co-funded under the European Regional Development Fund under Grant Number 13/RC/2077 and 13/RC/2077-P2.]}

\end{document}